\documentclass[review, 1p]{elsarticle}

\usepackage{lineno,hyperref}
\modulolinenumbers[5]

\journal{Journal of \LaTeX\ Templates}

\usepackage{url}
\usepackage[utf8]{inputenc}
\usepackage{tabularx}
\usepackage{orcidlink}
\usepackage{graphicx}% Include figure files
\usepackage{dcolumn}% Align table columns on decimal point
\usepackage{bm}% bold math
\usepackage[utf8]{inputenc}
\usepackage[T1]{fontenc}
\usepackage{mathptmx}
\usepackage{etoolbox}
\usepackage{xcolor}
\usepackage{graphicx}
\usepackage{lineno,hyperref}
\usepackage{blindtext}
\usepackage{epsfig}
\usepackage{latexsym}
\usepackage{graphics}
\usepackage{epsfig}
\usepackage{amsmath,amssymb,amsfonts,theorem,color,bm}
\usepackage{multirow, multicol, boldline}
\usepackage{booktabs}
\usepackage{stfloats} 
\usepackage{float}
\usepackage{subfig}
\usepackage{soul}
\usepackage{caption}
\hypersetup{colorlinks,allcolors=black}
\usepackage[normalem]{ulem}

\DeclareMathAlphabet\mathbfcal{OMS}{cmsy}{b}{n}

\begin{document}

\begin{frontmatter}

\title{A Critical Assessment of Pattern Comparisons Between POD and Autoencoders in Intraventricular Flows}

\newcommand{\orcidEL}{0000-0002-4514-6471}
\newcommand{\orcidJGM}{0000-0002-7422-5320}
\newcommand{\orcidSLCM}{0000-0003-3605-7351}
\newcommand{\orcidABN}{0000-0002-8539-5405}
\newcommand{\orcidPK}{0000-0001-8337-2122}

\author[addressUPM]{E. Lazpita \orcidlink{\orcidEL}}
\cortext[mycorrespondingauthor]{Corresponding author}
\ead{e.lazpita@upm.es}

\author[addressUPM]{A. Bell-Navas \orcidlink{\orcidABN}}
\ead{a.bell@upm.es}
\author[addressUPM,addressCSC]{J. Garicano-Mena \orcidlink{\orcidJGM}}
\ead{jesus.garicano.mena@upm.es}
\author[addressUH]{P. Koumoutsakos \orcidlink{\orcidPK}}
\ead{petros@seas.harvard.edu}
\author[addressUPM,addressCSC]{S. Le Clainche \orcidlink{\orcidSLCM}}
\ead{soledad.leclainche@upm.es}

\address[addressUPM]{ETSI Aeron\'autica y del Espacio - Universidad Polit\'ecnica de Madrid, 28040 Madrid, Spain}
\address[addressCSC]{Center for Computational Simulation (CCS), 28660 Boadilla del Monte, Spain}
\address[addressUH]{John A. Paulson School of Engineering and Applied Sciences, Harvard University, Boston, MA 02134, Estados Unidos}

\begin{abstract}

Understanding intraventricular hemodynamics requires compact and physically interpretable representations of the underlying flow structures, as characteristic flow patterns are closely associated with cardiovascular conditions and can support the early detection of cardiac deterioration. Conventional visualization of velocity or pressure fields, however, provides limited insight into the coherent mechanisms driving these dynamics. Reduced-order modeling techniques, such as Proper Orthogonal Decomposition (POD) and Autoencoder (AE) architectures, offer powerful alternatives for extracting dominant flow features from complex datasets. This study systematically compares POD with several AE variants (Linear, Nonlinear, Convolutional, and Variational) using left ventricular flow fields obtained from computational fluid dynamics simulations. We show that, for a suitably chosen latent dimension, AEs produce modes that become nearly orthogonal and qualitatively resemble POD modes that capture a given percentage of kinetic energy. As the number of latent modes increases, AE modes progressively lose orthogonality, leading to linear dependence, spatial redundancy, and the appearance of repeated modes with substantial high-frequency content. This degradation reduces interpretability and introduces noise-like components into AE-based reduced-order models, potentially complicating their integration with physics-based formulations or neural-network surrogates. The extent of interpretability loss varies across AE architectures, with nonlinear, convolutional, and variational models exhibiting distinct behaviors in orthogonality preservation and feature localization. Overall, the results indicate that Autoencoders can reproduce POD-like coherent structures under specific latent-space configurations, while highlighting the need for careful mode selection to ensure physically meaningful representations of cardiac flow dynamics.

\end{abstract}


\end{frontmatter}

%%%%%%%%%%%%%%%%%%%%%%%%%%%%%%%%%
%%%%%%%%%%%%%%%%%%%%%%%%%%%%%%%%%
\section{\label{sec:Introduction} Introduction}

The accurate analysis of intraventricular flow is fundamental to understand the mechanisms that govern cardiac function. The flow inside the left ventricle (LV) is characterized by strong three-dimensionality, temporal variability, and the presence of large-scale vortical structures that evolve throughout the cardiac cycle. Among these,
the incoming blood flow that fills the LV cavity during the diastole, and the posterior ejection through the aortic valve opening during systole play a crucial role, as this facilitates an efficient redirection of the inflow toward the outflow tract, minimizing energy losses and promoting effective ejection during systole. The formation, evolution, and eventual breakdown of this vortex ring are therefore considered key indicators of pumping efficiency and overall ventricular performance \cite{Lazpita2025Characterizing, Nagargoje2025Review}.

Cardiac flow data can be obtained from a wide range of sources, including medical imaging techniques such as magnetic resonance imaging (MRI) \cite{Markl2012Flow} or computed tomography (CT) \cite{Pontone2022Clinical, Niemeyer1995Nuclear}, experimental approaches such as particle image velocimetry (PIV) \cite{Kilner2000Asymmetric, Querzoli2010Effect} and in~vitro and in~silico models \cite{Korn2020Silico, Colebank2022Silico}, and computational fluid dynamics (CFD) simulations that numerically resolve the coupled fluid and structural dynamics of the ventricle \cite{Korte2023Hemodynamic, Zingaro2024Electromechanics}. Each of these modalities provides complementary information about the hemodynamic environment, typically expressed in terms of velocity fields, pressure distributions, wall shear stress distributions, or vorticity patterns. However, these visualization techniques only offer a qualitative perspective, and often fail to reveal the intrinsic mechanisms driving flow evolution, the characteristic frequencies associated with cardiac dynamics, and the onset of flow instabilities such as vortex breakdown \cite{Nagargoje2025Review, Begiashvili2023Data}.

To overcome these limitations, modal decomposition techniques have gained increasing attention. Proper Orthogonal Decomposition (POD) \cite{Lumley1967Structure}, Dynamic Mode Decomposition (DMD) \cite{Schmid2010Dynamic} and its robust extension Higher Order DMD (HODMD) \cite{LeClainche2017Higher} allow a systematic extraction of coherent structures and dominant temporal patterns from complex flow fields. These methods have been successfully applied in various biomedical contexts, including arterial and venous flows \cite{Kazemi2022Reduced, Habibi2020Data}. However, applications to the heart cavity dynamics remain relatively limited, although some studies are starting to arise. For instance, Ref. \cite{Wu2023Flow} combined the Shake-the-Box algorithm—an advanced Lagrangian particle tracking method—with POD to analyze the three-dimensional and time-resolved nature of intraventricular flow. Similarly, Ref. \cite{Borja2024Deriving} employed POD in patient-specific simulations, identifying characteristic flow patterns and quantifying the energy contribution of each mode. 

Recent research has extended these methods to incorporate temporal dynamics through DMD. Studies such as Ref. \cite{Borja2016Dynamic} and Ref. \cite{Mikhail2017Proper} extracted coherent modes associated with specific frequencies and proposed modal biomarkers to characterize cardiac performance and pathology. In a subsequent contribution, Ref. \cite{DiLabbio2019reduced} combined POD and DMD to construct a reduced-order model of left ventricular flow under pathological conditions. These efforts demonstrate the potential of modal decomposition techniques to reveal the underlying physics of intraventricular hemodynamics and to develop compact representations of complex flow phenomena.

In parallel, the advent of deep learning has introduced a new paradigm for analyzing high-dimensional biomedical and fluid dynamics data. In fluid mechanics, Convolutional and Autoencoder-based neural networks have been used to extract discriminating features represented in low-dimensional latent spaces, reconstruct flow fields, and predict their temporal evolution with remarkable accuracy \cite{Eivazi2022Towards}. We have used POD/DMD as data-driven approaches alternative to traditional reduced-order models, enabling the identification of nonlinear dynamics and subtle spatial correlations that linear methods may overlook.

In the biomedical domain, deep learning has shown great promise for the analysis of cardiac imaging and hemodynamic data. For example, Ref.~\cite{Groun2022Higher} applied HODMD to analyze echocardiography imaging and identify features associated with heart diseases. The resulting dynamic descriptors were subsequently integrated with deep learning frameworks for automatic heart disease recognition, achieving significantly higher accuracy than conventional image-based models \cite{Bell2025Automatic}. These developments highlight the potential of combining modal decomposition and deep learning to uncover physiologically meaningful flow patterns and to enhance diagnostic and predictive tools in cardiovascular research.

Building upon these advances, the present work explores data-driven methodologies for extracting, analyzing, and interpreting coherent flow structures in intraventricular CFD data. By bridging physics-based modal decomposition techniques with deep learning approaches, this study aims to deepen the understanding of the dynamic mechanisms governing left ventricular flow and to establish a foundation for efficient, interpretable, and predictive models of cardiac hemodynamics.

In particular, this work addresses a gap in the current literature regarding the interpretability of Autoencoder-based modes when applied to cardiac flow fields. While Autoencoders have demonstrated remarkable capabilities for nonlinear dimensionality reduction, their ability to generate modal structures comparable to those produced by POD remains insufficiently understood. Here, we investigate under which conditions AE latent representations yield coherent structures that qualitatively resemble POD modes, and how this relationship evolves as the number of latent dimensions increases. A key aspect of this analysis is the observation that AE modes may lose orthogonality when the latent space grows, leading to spatial redundancy, mode repetition, and the appearance of high-frequency content that behaves similarly to noise. Such effects can obscure the underlying flow physics and raise important questions about the reliability of AE-based reduced-order models.

Understanding these behaviors is essential for the future development of physics-aware data-driven models. Non-orthogonal and redundant AE modes can hinder the enforcement of physical constraints, limit the interpretability of learned representations, and introduce instability when coupling the reduced-order model with neural-network surrogates or hybrid physics–ML frameworks. By clarifying the circumstances under which AE architectures produce physically meaningful modal structures—and when they do not—this study provides a systematic foundation for designing reduced-order strategies that preserve relevant cardiac flow features while avoiding the inclusion of noise-like or non-reproducible dynamics. The findings presented here thus contribute novel insights into the integration of deep learning with physics-based decomposition methods, offering guidance for the development of future predictive, clinically oriented, and computationally efficient models of intraventricular hemodynamics.

The article is organized as follows. In Section~\ref{sec:Methods}, the modal decomposition framework employed in this study is described, outlining the methodologies used for feature extraction through POD and AEs. Section~\ref{sec:Database} introduces the database employed for model training and validation, describing the numerical setup and preprocessing procedures. The comparative results between POD and AE representations are presented and discussed in Section~\ref{sec:Results}, focusing on the capability of each method to capture coherent structures and flow dynamics. Finally, the main findings and conclusions are summarized in Section~\ref{sec:Conclusions}.

%%%%%%%%%%%%%%%%%%%%%%%%%%%%%%%%%
%%%%%%%%%%%%%%%%%%%%%%%%%%%%%%%%%
\section{\label{sec:Methods} Foundations of POD and Autoencoder-Based Decomposition}

To enable the application of data-driven reduction methods, the raw CFD outputs must be reformatted appropriately. The anatomical variability and temporal evolution of the ventricular geometries considered requires the construction of a fixed spatial frame. Specifically, an structured Cartesian grid is defined such that it encloses the ventricle throughout the entire cardiac cycle. Simulation results are interpolated onto this reference mesh, producing a spatiotemporal representation that includes both regions occupied by fluid and those left empty during the motion. The resulting dataset can be expressed as a five-dimensional tensor \(\mathbf{A}\) of shape \( (J_1 \times J_2 \times J_3 \times J_4 \times K) \), where \( J_1 \) indicates the number of physical quantities (e.g., velocity components, pressure), \( J_2, J_3, J_4 \) define the spatial discretization in three dimensions of the domain, and \( K \) is the total number of temporal instances/samples.

In this study, we adopt two complementary strategies to construct reduced representations of the intraventricular flow data. On the one hand, we employ data-driven modal decomposition techniques, specifically Proper Orthogonal Decomposition (POD), which provide interpretable, energy-ranked modes based on linear algebraic principles. On the other hand, we explore nonlinear dimensionality reduction through deep learning, utilizing AEs architectures trained to extract compact latent spaces. Both approaches aim to reveal the dominant spatiotemporal patterns within the dataset, improve reconstruction, and to shed light on the underlying flow dynamics.

\subsection{\label{subsec:pod} Proper Orthogonal Decomposition}

To apply the POD method, the tensor is first reshaped into a matrix \( \mathbf{V} \in \mathbb{R}^{J \times K} \), where \( J = J_1 \cdot J_2 \cdot J_3 \cdot J_4 \). In this format, each column corresponds to a single temporal snapshot of the full domain, while rows index spatial locations and physical variables.

POD \cite{Berkooz1993Proper} is a classical approach for extracting dominant spatial structures in high-dimensional datasets. It provides a low-rank approximation that optimally captures the variance of the system. In our study, the decomposition is achieved via the Singular Value Decomposition (SVD) \cite{Sirovich1987Turbulence}, yielding

\begin{equation}
    \mathbf{V} = \mathbf{U} \boldsymbol{\Sigma} \mathbf{T}^T = \sum_{n=1}^{\min{\left(J, K\right)}} s_n \mathbf{u}_n \cdot \mathbf{t}_n^T,
    \label{eq:SVD}
\end{equation}
where \( \mathbf{U} \in \mathbb{R}^{J \times N} \) contains the spatial POD modes, \( \boldsymbol{\Sigma} \in \mathbb{R}^{N \times N} \) is a diagonal matrix with singular values \( s_1 \geq s_2 \geq \dots \geq s_N \), \( \mathbf{T} \in \mathbb{R}^{K \times N} \) holds the temporal coefficients, and \( N \leq \min{\left(J, K \right)} \) is the rank for the best approximation of \( \mathbf{V} \) according to Eckart–Young–Mirsky theorem. The matrix transpose is denoted by \( ()^T \).

To control the dimensionality of the reduced representation, a tolerance \( \varepsilon \) is defined such that only the first \( r  \leq N \) modes are retained, ensuring \( s_{r + 1} / s_1 \leq \varepsilon \). Truncating the decomposition in this way provides a compact reconstruction of the dataset while filtering out low-energy fluctuations and numerical noise.

The reduced-order reconstruction can be expressed by limiting the summation in Eq.~\eqref{eq:SVD} to the first \( r \) terms. The spatial basis \( \mathbf{U} \) obtained through this process reveals the most energetic flow features, offering a physically interpretable basis for analysis.

The singular values \( s_n \) are associated with the energy content of each mode. To quantify their relative contribution, the normalized modal energy is computed as

\begin{equation}
    e_n = \frac{s_n^2}{\sum_{n=1}^N s_n^2} \cdot 100 \, ,
    \label{eq:energy_svd}
\end{equation}
providing a measure of the proportion of the total energy captured by the \( n \)-th mode.

\subsection{\label{subsec:ae} Autoencoder Architectures}

To complement the linear decomposition approach of POD, we investigate the use of Autoencoders (AEs) as nonlinear data-driven models for reduced-order representation. These architectures are designed to learn compact encodings of high-dimensional inputs by minimizing the reconstruction error through a bottleneck structure.

The Autoencoder comprises two main components: an encoder \( \mathcal{E}_{\pmb{\theta}} \) and a decoder \( \mathcal{D}_{\pmb{\phi}} \), parametrized by trainable weights \( \pmb{\theta} \) and \( \pmb{\phi} \), respectively. The encoder maps the input field \( \mathcal{X}_k \in \mathbb{R}^{J_1 \times J_2 \times J_3 \times J_4} \) to a low-dimensional latent vector \( \mathbf{z}_k \in \mathbb{R}^r \), where \( r \ll J \). The decoder reconstructs the field from this latent representation, yielding an output \( \widehat{\mathcal{X}}_k = \mathcal{D}_{\boldsymbol{\pmb{\phi}}}(\mathbf{z}_k) \). The network is trained by minimizing the reconstruction loss over all \( K \) snapshots:

\begin{equation}
    \mathcal{L}_{\text{MSE}}(\pmb{\theta}, \pmb{\phi}) = \frac{1}{K} \sum_{k=1}^{K} \left\| \mathcal{X}_k - \mathcal{D}_{\pmb{\phi}} \left( \mathcal{E}_{\pmb{\theta}}(\mathcal{X}_k) \right) \right\|_2^2 \, .
    \label{eq:ae_loss}
\end{equation}

To allow a fair comparison with POD, the latent dimension \( r \) is selected to match the number of retained modes in the reduced-order linear model. All models are trained with batch normalization and early stopping to improve convergence and mitigate overfitting. Architectural hyperparameters, such as the number of layers, filter size, stride, and activation function, are fixed across simulations for consistency.

Depending on the case, the network is constructed using either fully connected layers or a hybrid architecture combining convolutional and dense layers, with the aim of capturing spatial features more effectively. This design strategy follows recent studies demonstrating the robustness of Convolutional Autoencoders (CAEs) for complex fluid flow fields \cite{Eivazi2022Towards}.

In addition to standard AEs, we also consider the Variational Autoencoder (VAE) \cite{kingma2013auto} architecture, which introduces a probabilistic formulation of the latent space. Instead of encoding a single deterministic vector, the encoder predicts a continuous Gaussian distribution for each snapshot, parameterized by its mean \( \pmb{\mu}_k \) and standard deviation \( \pmb{\sigma}_k \). A latent sample \( \mathbf{z}_k \) is then drawn from that Gaussian distribution \( \mathcal{N}(\pmb{\mu}_k, \text{diag}(\boldsymbol{\pmb{\sigma}}_k^2)) \), enabling the model to generalize over a continuous distribution of latent codes rather than specific discrete encodings.

In this case, the total loss function includes both the reconstruction error and a regularization term that penalizes deviations from the unit Gaussian prior via the Kullback-Leibler (KL) divergence \cite{kingma2013auto}:

\begin{equation}
    \mathcal{L}_{\text{VAE}}(\pmb{\theta}, \pmb{\phi}) = \mathcal{L}_{\text{MSE}} + \beta \cdot \mathcal{L}_{\text{KL}} = \frac{1}{K} \sum_{k=1}^{K} \left[ \left\| \mathcal{X}_k - \widehat{\mathcal{X}}_k \right\|^2 + \beta \cdot D_{\text{KL}}\left( \mathcal{N}(\pmb{\mu}_k, \text{diag}(\pmb{\sigma}_k^2)) \, \| \, \mathcal{N}(\mathbf{0}, \mathbf{I}) \right) \right] \, .
    \label{eq:vae_loss}
\end{equation}

The weighting parameter \( \beta \) controls the strength of the regularization and balances the trade-off between reconstruction fidelity and latent space continuity.

\begin{figure}[h]
    \centering
    \includegraphics[trim = 0 0 0 0, clip, width=\textwidth]{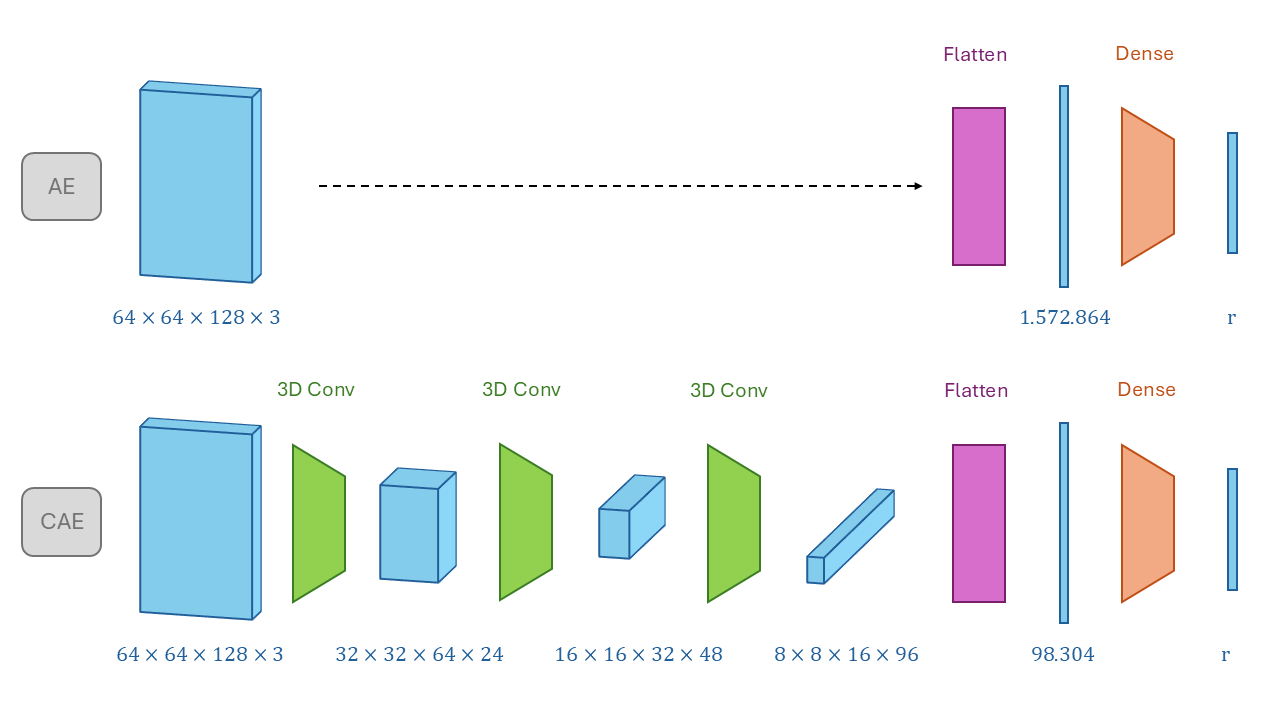}
    \caption{Schematic illustration of the two main different Autoencoder (AE) architectures considered: (Top) the fully connected and (bottom) the Convolutional Autoencoders (CAEs). All layers in each configuration have identical activation function listed in Tab.~\ref{tab:cases}.}
    \label{fig:architectures}
\end{figure}

The different Autoencoder configurations investigated in this work are summarized in Tab.~\ref{tab:cases}, which reports the main architectural components and hyperparameters for each case: Linear Autoencoder (LIN), Non-linear Autoencoder (non-LIN), Convolutional Autoencoder (CAE) and Variational Autoencoder (VAE). These configurations were selected based on prior studies that employed them in fluid dynamics to identify flow patterns \cite{Eivazi2022Towards} and even to predict future flow evolution \cite{Abadia2022Predictive, Abadia2025Generalization}. For clarity, Fig.~\ref{fig:architectures} depicts the two principal architectures considered: a Fully Connected Autoencoder and a Convolutional Autoencoder (CAE). The Variational Autoencoder is not illustrated, as its structure is identical to the convolutional model except for the inclusion of a latent sampling layer that defines the variational space. The latent dimension \( r \) is fixed across all models to ensure a consistent basis for the pattern comparison presented in Section~\ref{sec:Results}.

\begin{table}[H]
\centering
\begin{tabular}{| l | c | c | c | c | c | c |}
\hline
\textbf{Case} & \textbf{Conv. Layers} & \textbf{Dense Layers} & \textbf{Activation Function} & \( \beta \) & Batch Size & Epochs \\
\hline
LIN & 0 & 1 & Linear & - & 50 & 500 \\
Non-LIN & 0 & 1 & Leaky-ReLU & - & 50 & 500 \\
CAE & 3 & 1 & Leaky-ReLU & - & 50 & 500 \\
VAE & 3 & 1 & Leaky-ReLU & 0.1 & 50 & 500 \\
\hline
\end{tabular}
\caption{Autoencoder architectures used for dimensionality reduction. The table lists the number of convolutional and dense layers, the activation function, the KL regularization coefficient \( \beta \), the batch size and the number of epochs used in each case. The size of each layer is defined in Fig.~\ref{fig:architectures}.}
\label{tab:cases}
\end{table}

Once trained, these Autoencoders provide a nonlinear mapping to a low-dimensional latent space that can be used for visualization, clustering, or reconstruction. While the latent vectors lack the orthonormality and energy ranking of POD modes, they can capture nonlinear relationships and spatially coherent structures that remain hidden to linear techniques. % The energy associated with each latent vector is defined in a manner similar to the POD formulation and is quantified here using the squared norm of the corresponding latent vector, \( ||z_k|| \).

\subsection{\label{subsec:metrics} Metrics}

The accuracy of the reconstructed flow fields is quantified using the Relative Root Mean Square Error (RRMSE). For a set of $r$ retained modes, the RRMSE is defined as
\begin{equation}
\text{RRMSE (\%)} = \frac{\sqrt{ \| \mathbf{A} - \mathbf{A}_{\text{rec}} \|^2}}{\sqrt{ \| \mathbf{A} \|^2}} \cdot 100,
\label{eq:rrmse}
\end{equation}
where $\mathbf{A}$ is the original data, and $\mathbf{A}_{\text{rec}}$ its reconstruction using the retained modes. This metric provides a normalized measure of the reconstruction error, with lower values indicating better agreement between the reduced-order representation and the original data.

To assess the independence of the modes, we define a global orthogonality measure based on the Modal Assurance Criterion (MAC) matrix. Let $\text{MAC}_{ij}$ denote the MAC between modes $i$ and $j$, and let $r$ be the selected latent space size. The orthogonality measure is defined as
\begin{equation}
\mathbb{O} = 1 - \frac{\| \text{MAC} - I \|_F}{\sqrt{r(r-1)}},
\label{eq:orthogonality}
\end{equation}
where $\| \cdot \|_F$ is the Frobenius norm, and $I$ is the identity matrix. This metric ranges from 0 to 1, with $O=1$ corresponding to a set of perfectly orthogonal modes. It captures the overall similarity among modes and complements the RRMSE, ensuring that the retained modes provide physically independent and non-redundant contributions to the reduced-order representation.

%%%%%%%%%%%%%%%%%%%%%%%%%%
%%%%%%%%%%%%%%%%%%%%%%%%%%
\section{\label{sec:Database} Intraventricular Flow Simulation Dataset}

To assess and compare the performance of the Autoencoder architectures introduced in previous sections against the POD method, we employ a simulation database based on an idealized model of the left ventricle. This simplified geometry, originally proposed in Ref.~\cite{Zheng2012Computational}, consists of a semi-ellipsoidal chamber designed to reproduce essential features of ventricular anatomy while maintaining geometric regularity, thereby facilitating the application of reduced-order modeling techniques.

To ensure that the model exhibits physiologically realistic characteristics, especially regarding the minimum volume reached at end-systole, a set of dimensionless geometric parameters is defined. These parameters, presented in Tab.~\ref{tab:geometry_parameters}, are scaled with respect to the base radius \( a \) of the ventricular cavity, which is taken as the characteristic length.

\begin{table}[H]
\centering
\begin{tabular}{| l | c |}
\hline
\textbf{Parameter} & \textbf{Value}  \\
\hline
Chamber length \( b/a \)          & 4.0 \\
Inlet diameter \( D/a \)          & 1.2 \\
Outlet diameter \( d/a \)         & 0.4 \\
Tube height \( H/a \)             & 2.4 \\
Inlet offset \( C/a \)            & 0.275 \\
Outlet offset \( c/a \)           & 0.675 \\
Reference radius \( a \) [cm]     & 2.0 \\
\hline
\end{tabular}
\caption{Dimensionless parameters defining the geometry of the ventricular model, with base radius \( a \) used as the reference length (cfr. Fig.~\ref{fig:geometry}).}
\label{tab:geometry_parameters}
\end{table}

\begin{figure}[h]
    \centering
    \includegraphics[trim = 0 0 130 0, clip, width=\textwidth]{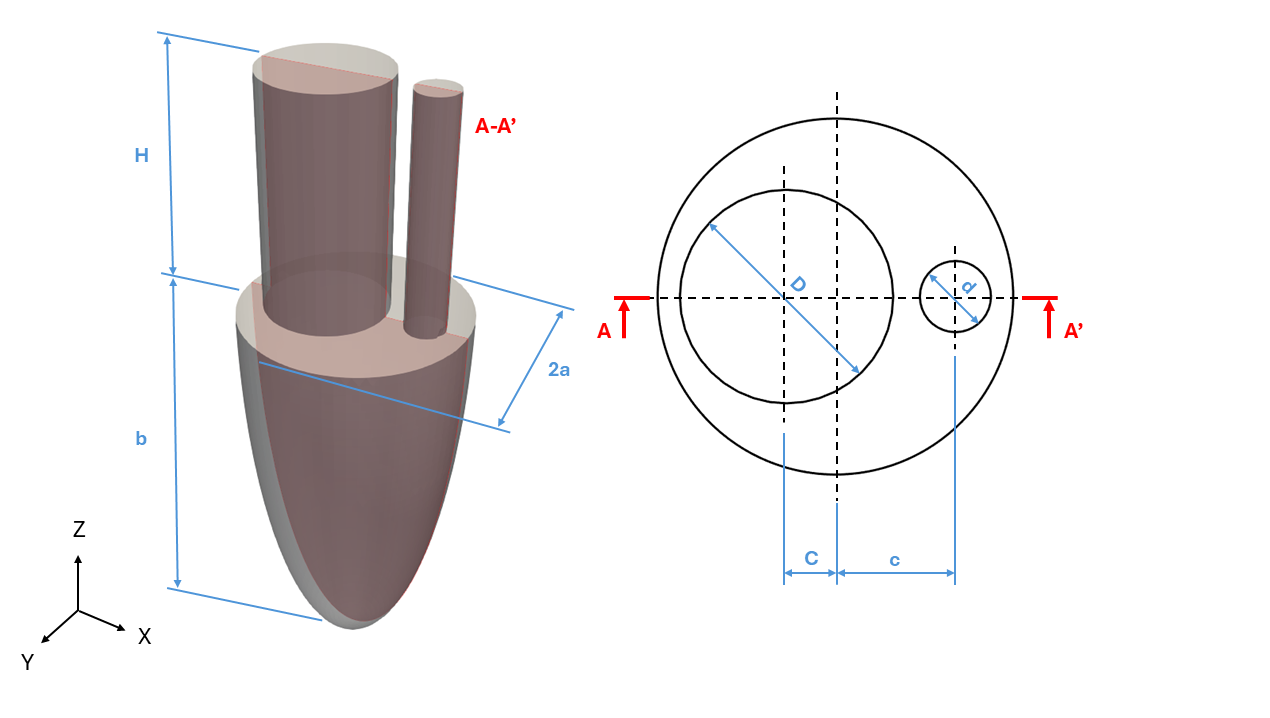}
    \caption{Schematic of the idealized left ventricle model. The red area highlights the symmetry plane A-A' employed in post-processing.}
    \label{fig:geometry}
\end{figure}

Fig.~\ref{fig:geometry} illustrates the configuration of the left ventricle model used in this work, including the main dimensions and orientation. The vertical symmetry plane \( y = 0 \), consistently used in the visualization of the results, is marked in red.

The fluid dynamics within the chamber is simulated using the \textit{Ansys Fluent} solver~\cite{Fluent}. Blood is represented as an incompressible Newtonian fluid with density \(\rho = 1060\, \text{kg}\,\text{m}^{-3}\) and dynamic viscosity \(\mu = 0.004\, \text{Pa}\cdot\text{s}\), following Ref. \cite{Korte2023Hemodynamic}. The simulations are carried out under conditions representative of a cardiac cycle, with dynamic boundaries that reflect the opening and closing phases of the ventricular valves. During diastole, inflow is enabled and the outlet boundary is sealed, while the configuration is reversed during systole. The motion of the ventricular wall is imposed through a series of meshes that evolve in time and induce pressure gradients capable of driving the fluid motion.

During the simulation, Reynolds numbers \( \mathrm{Re} \) of up to 5500 are attained. This value is computed using the peak inlet velocity and the characteristic inlet diameter. Although this range suggests transitional or even turbulent conditions, the flow is assumed laminar, following the rationale adopted in other computational studies~\cite{He2022Numerical, Tagliabue2017Complex}. This assumption allows for the accurate resolution of coherent vortical structures, including the vortex ring generated during early filling.

A complete description of the simulation protocol, modeling assumptions, and computational setup used to generate the databases analyzed in this work is provided in our previous studies~\cite{Lazpita2024ECCOMAS, Lazpita2024Modeling, Lazpita2025Characterizing}. Simulations were executed on the \textit{Magerit} high-performance computing platform hosted at CeSViMa ("Centro de Supercomputación y Visualización de Madrid"), utilizing 40 message-passing interface processes on a single compute node.

The Q-criterion \cite{Hunt1988Turbulent} is a scalar measure commonly used to detect regions of dominant rotation within a flow. It relies on the balance between the rotational and extensional components of the velocity gradient tensor, and is expressed as

\begin{equation}
Q = \frac{1}{2} \left( |\bm{\Omega}|^2 - |\mathbf{S}|^2 \right),
\end{equation}
where $\mathbf{S} = \tfrac{1}{2}\left( \nabla \mathbf{u} + (\nabla \mathbf{u})^\top \right)$ denotes the symmetric component of the velocity gradient (strain-rate tensor), and $\bm{\Omega} = \tfrac{1}{2}\left( \nabla \mathbf{u} - (\nabla \mathbf{u})^\top \right)$ represents its antisymmetric part (vorticity tensor). Regions with $Q > 0$ correspond to areas where rotation exceeds strain, making the criterion particularly useful for visualizing coherent vortical features such as the vortex ring.

Time is non-dimensionalized with respect to the period \( T \) of a complete cardiac cycle, such that \( t^* = t/T \). This facilitates comparisons between different phases of the simulation. Diastole occupies the interval \( t^* \in [i,\, i + 0.67] \), while systole spans \( t^* \in [i + 0.67,\, i + 1] \), with \( i \) denoting the index of the cardiac cycle, $i=[10, 11, \cdots, 19]$, with $i \in \mathbb{N}$.

A central feature of intraventricular flow is the generation of a vortex ring during diastole due to mitral inflow~\cite{Le2012Vortex, Toger2012Vortex, Nagargoje2025Review}. This structure undergoes deformation and eventual breakup, which are known to correlate with ventricular efficiency. Capturing its development and breakdown accurately is essential for meaningful pattern recognition.

\begin{figure}[h]
    \centering
    \includegraphics[trim = 0 100 0 0, clip, width=\textwidth]{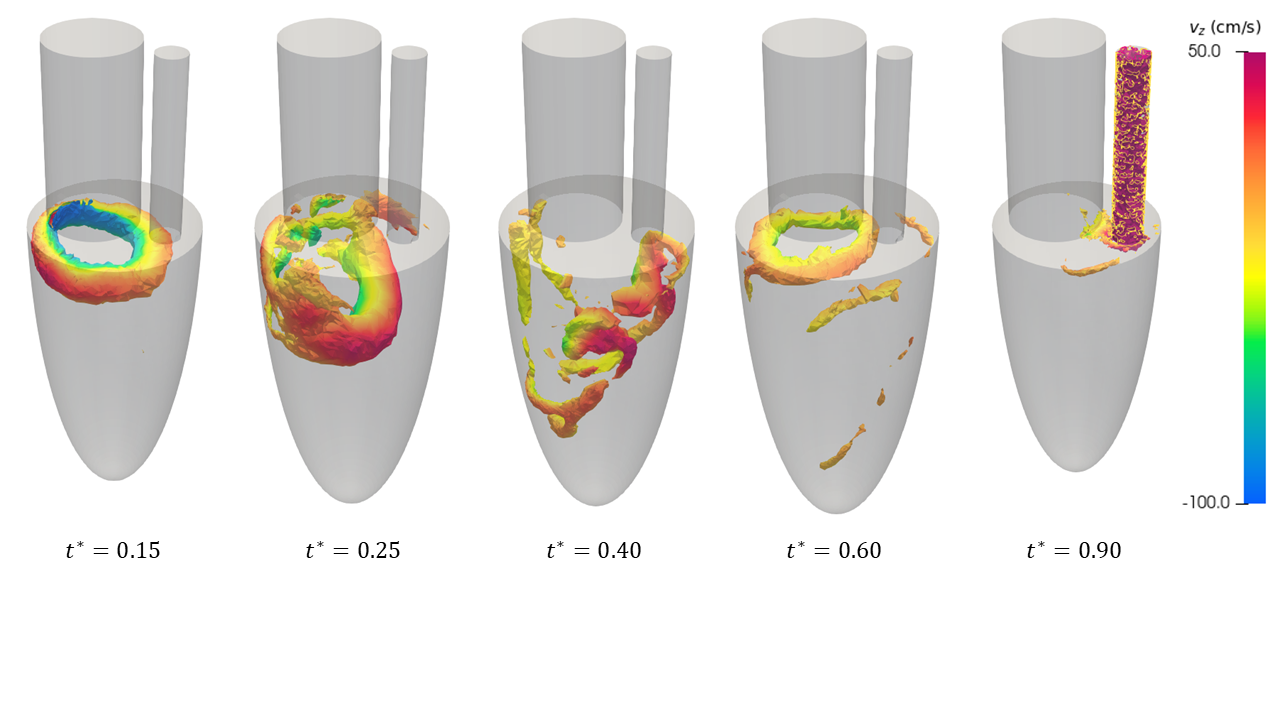}
    \caption{Flow visualization at different stages of the cardiac cycle. Green surfaces show the Q-criterion at a value of 2500. Background: vorticity field on the midplane, in the range \( [-100, 100]\, \text{s}^{-1} \).}
    \label{fig:database}
\end{figure}

Snapshots of the flow field at selected time instants are shown in Fig.~\ref{fig:database}. Vortical structures are visualized using a fixed value of Q-criterion; those structures are colored by the out-of-plane vorticity in the A-A' symmetry plane. It is evident that the vortex ring in this configuration dissipates rapidly, beginning to destabilize by \( t^* = 0.25 \) and vanishing before \( t^* = 0.40 \).

The dataset of our complete simulation contains 10 cycles of simulation with initial cycle having transitory noise. However, in this study, we collect data directly from cycle 10 onwards, where the transient noise is already mitigated and convergence of transient dynamics can be ensured ~(see Ref.~\cite{Lazpita2025Efficient}). Velocity fields are sampled at intervals of \( \Delta t^* = 0.05 \), resulting in 20 snapshots per cycle and a total of \( K = 200 \) samples. The reduced-order models are trained using the first 6 cycles of this dataset, while the subsequent 4 cycles are reserved for validation and testing.

Tab.~\ref{tab:partition} summarizes the partitioning strategy. Although prior works often limit simulations to 6 cycles~\cite{Cane2022CFD, Grunwald2022Intraventricular, Korte2023Hemodynamic}, our observations suggest that such durations may not suffice to capture discriminating patterns in complex flow regimes. Hence, the present database spans a longer temporal window to improve robustness in pattern extraction.

\begin{table}[H]
\centering
\begin{tabular}{| l | c | c | c |}
\hline
\textbf{Set} & \textbf{Cycles} & \textbf{Time Interval \( t^* \)} & \textbf{Snapshots \( K \)} \\
\hline
Training    & 6   & [10, 16]    & 120 \\
Validation  & 2   & [16, 18]    & 40  \\
Testing     & 2   & [18, 20]    & 40  \\
\hline
\end{tabular}
\caption{Summary of the dataset segmentation by cardiac cycles, normalized time window, and number of snapshots. The sampling of the cardiac data was started after 10 periods.}
\label{tab:partition}
\end{table}

%%%%%%%%%%%%%%%%%%%%%%%%%%%%%%%%%
%%%%%%%%%%%%%%%%%%%%%%%%%%%%%%%%%
\section{\label{sec:Results} Evaluation of Modal Structures and Reduced-Order Performance}

The analysis focuses on the ability of POD and Autoencoders to capture the main features of the left ventricle model flow. Both methods provide reduced representations of the velocity field, but they differ in how the reduced bases are constructed and in the degree of physical interpretability they offer. In POD, the basis functions are ranked by their energy content, while in Autoencoders, this ordering depends on the chosen latent space dimension. This section presents the results obtained with each case listed before in Tab. \ref{tab:cases} for different latent dimensions $r$ and compares their performance in terms of energy distribution, temporal dynamics, and interpretability of the reduced modes.

\begin{table}[h]
\centering
\setlength{\tabcolsep}{6pt} % widen columns
\begin{tabular}{|c|c|c|c|c|c|c|c|c|c|c|}
\hline
\multirow{2}{6pt}{$r$} & \multicolumn{2}{c|}{POD} & \multicolumn{2}{c|}{LIN} & \multicolumn{2}{c|}{Non-LIN} & \multicolumn{2}{c|}{CAE} & \multicolumn{2}{c|}{VAE} \\
\cline{2-11}
& RRMSE & $\mathbb{O}$ & RRMSE & $\mathbb{O}$ & RRMSE & $\mathbb{O}$ & RRMSE & $\mathbb{O}$ & RRMSE & $\mathbb{O}$ \\
\hline
5  & 27.2\% & 1 & 31.3\% & 0.46 & 35.0\% & 0.13 & 25.3\% & 0.71 & 25.3\% & 0.87\\
10 & 17.7\% & 1 & 25.6\% & 0.54 & 32.9\% & 0.09 & 24.5\% & 0.47 & 25.1\% & 0.73 \\
20 & 9.4\%  & 1 & 25.1\% & 0.21 & 26.6\% & 0.08 & 23.5\% & 0.43 & 25.2\% & 0.65 \\
40 & 0.5\%  & 1 & 35.3\% & 0.33 & 24.8\% & 0.08 & 22.3\% & 0.46 & 25.0\% & 0.62 \\
\hline
\end{tabular}
\caption{Reconstruction error (RRMSE) and orthogonality measure ($\mathbb{O}$) for different latent dimensions $r$ and model configurations.}
\label{tab:RRMSE}
\end{table}

Table~\ref{tab:RRMSE} presents the reconstruction errors (RRMSE, Eq.~\ref{eq:rrmse}) alongside the corresponding orthogonality measures ($\mathbb{O}$, Eq.~\ref{eq:orthogonality}) for different latent-space sizes $r$ and model configurations. As can be seen, the reconstruction accuracy shows limited sensitivity to the latent space dimension in the models with convolutional layers (CAE and VAE), which remain around 25\% across all cases. In contrast, the POD error decreases markedly as $r$ increases, reflecting its purely linear nature. This suggests that, while enlarging the latent space slightly improves accuracy, it does not substantially affect the nonlinear models’ performance. Instead, the choice of $r$ primarily affects the physical interpretability of the latent representations rather than the reconstruction accuracy, as reflected by the orthogonality metric, which tends to decrease with increasing latent-space size.

\begin{figure}[h]
    \centering
    \includegraphics[trim = 0 0 0 0, clip, width=\textwidth]{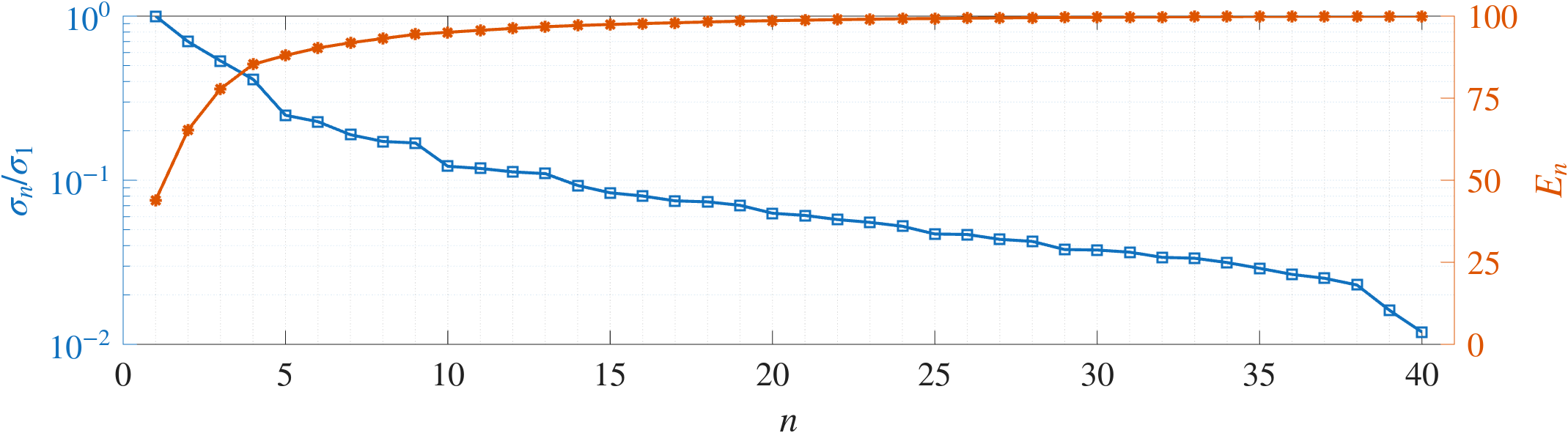}
    \caption{Decay of the normalized singular values (blue, left axis) and cumulative energy content $E_n = \sum_{k=1}^{n} e_k$ associated with Eq.~\ref{eq:energy_svd} (orange, right axis) for the POD modes.}
    \label{fig:energy-pod}
\end{figure}

The singular value distribution shows that, once normalized with respect to the leading value, the first five modes already exhibit a drop of about one order of magnitude, capturing approximately 88\% of the total energy content before the spectrum begins to saturate (Fig.~\ref{fig:energy-pod}). 

\begin{figure}[h]
    \centering
    \includegraphics[trim = 0 0 0 0, clip, width=\textwidth]{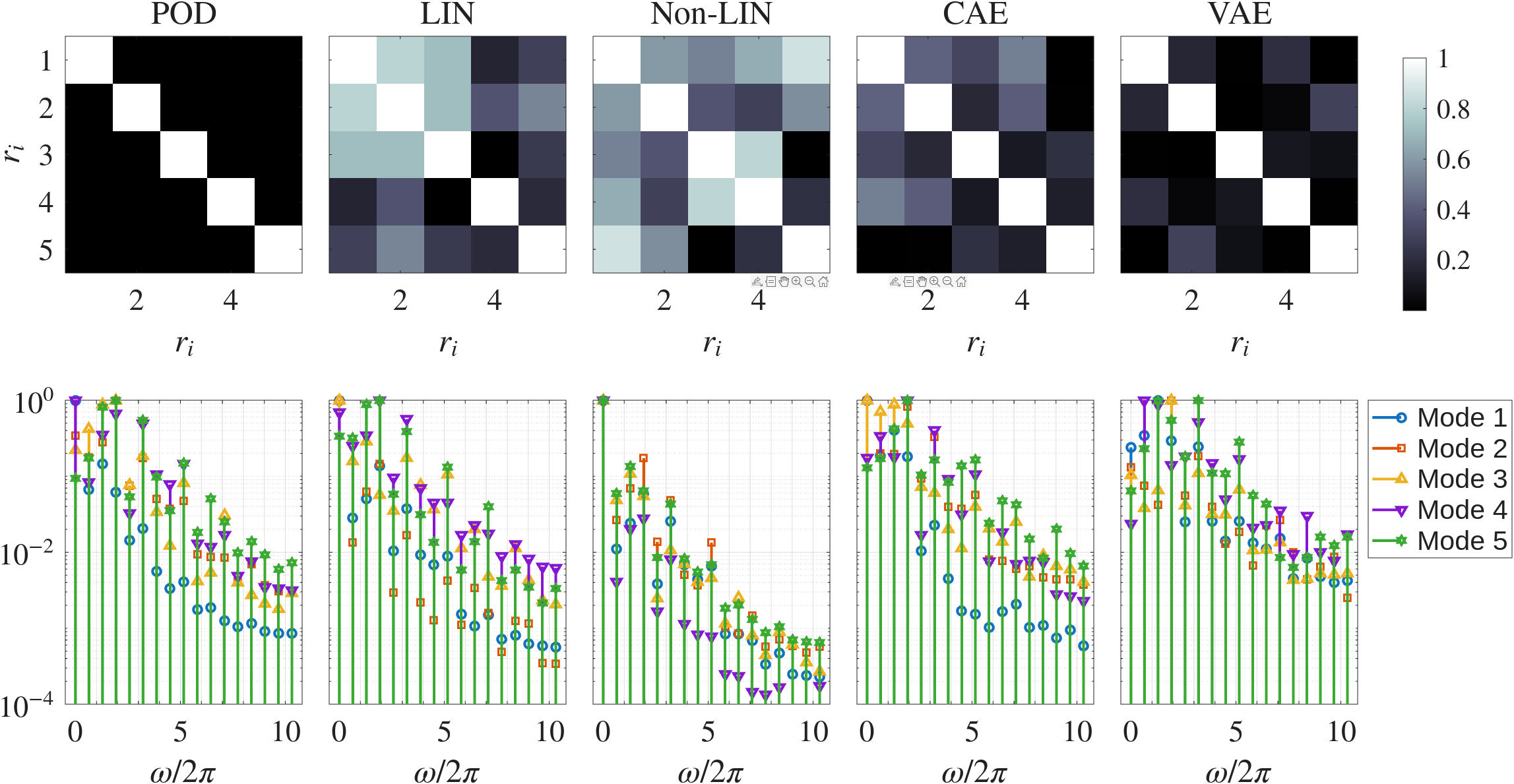}
    \caption{(Top) Correlation matrices and (bottom) normalized frequency spectra of the modes obtained for each case listed in Tab.~\ref{tab:cases}, using a latent space dimension of $r=5$ corresponding to a cumulative energy of $E_{5} = 88\%$. In the correlation matrices, black indicates complete orthogonality and white denotes full alignment between modes. The reconstruction errors associated with each case are reported in Tab.~\ref{tab:RRMSE}.}
    \label{fig:CorrelationMatrixFFT_r5}
\end{figure}

Fig.~\ref{fig:CorrelationMatrixFFT_r5} presents the correlation matrices and power spectral densities of the modes obtained for each case listed in Tab.~\ref{tab:cases} with an encoding dimension of $r=5$. The correlation matrices quantify the degree of orthogonality among the modes, whereas the spectral analysis of the associated chronos using FFT reveals their dominant frequency content. Modes with an orthogonality coefficient above 0.8 are considered mutually independent. The threshold of 0.8 is selected as a practical criterion to ensure that the correlation between two modes remains sufficiently low for them to be treated as distinct dynamical structures. An orthogonality (or correlation) coefficient below 0.2 in magnitude corresponds to over 80\% of the variance being unshared between the two modes, which indicates that the overlap in their spatial or temporal content is limited. In modal decomposition analyses, it is common to adopt such empirical thresholds to distinguish physically meaningful, independent structures from redundant or linearly dependent ones (see Ref.~\cite{Lazpita2022Generation}. Therefore, modes with an orthogonality coefficient above 0.8 are considered mutually independent, as they capture largely distinct flow features and contribute unique information to the reduced-order representation.

For the POD case, prior work ~\cite{Lazpita2025Characterizing} has shown that the first mode corresponds to a zero-frequency component representing the mean flow. The second and third modes display a dominant oscillation frequency, whereas the fourth and fifth modes contain contributions from both the fundamental frequency and its first harmonic. Together, these five modes reconstruct the coherent flow dynamics completing the set of structures that reconstruct the coherent flow behavior linked to vortex ring formation during early diastole. Owing to their mutual orthogonality, these modes represent independent flow mechanisms and provide a compact, physically interpretable description of the intraventricular flow.

This modal separation illustrates a fundamental feature of POD: the basis functions are ranked according to their energy content. As a result, the most energetic flow structures naturally align with dominant temporal scales, without the need to adjust the latent dimension manually. In contrast, AEs require an explicit choice of latent dimension, which strongly influences whether the extracted modes retain physical interpretability. A direct comparison between the two approaches is therefore necessary to assess how different decomposition strategies capture the underlying flow physics.

In contrast to POD, the modes extracted by AEs are not constrained to be orthogonal. The degree of orthogonality strongly depends on the network architecture, which in turn influences the extent to which the reduced modes retain physical interpretability. For the VAE, the extracted modes exhibit frequency content similar to that of POD, with minor variations depending on the mode. Most modes remain nearly orthogonal, and those that are not still present a high degree of linear independence. In this context, orthogonality implies that the modes are linearly independent, ensuring that their combination provides an optimal representation of the system dynamics.  

This property is progressively lost in other architectures. CAEs preserve a moderate degree of orthogonality, Linear Autoencoders (LIN) show only limited independence among modes, and Nonlinear Autoencoders (non-LIN) almost completely lack orthogonality. The frequency spectra reflect this trend. In CAEs (Fig. \ref{fig:CorrelationMatrixFFT_r5}, fourth column), many modes capture the dominant frequency mixed with its harmonics. In LIN (Fig. \ref{fig:CorrelationMatrixFFT_r5}, second column), two modes are dominated by the zero-frequency component and another contains only a weak oscillatory contribution, with non-zero frequency amplitudes that are substantially smaller than the zero-frequency content. In non-LIN (Fig. \ref{fig:CorrelationMatrixFFT_r5}, third column), all modes collapse into representations of the mean flow, with only faint traces of the dominant frequency.  

When orthogonality is absent, several modes become linearly dependent, effectively repeating the same structure in different forms. As a consequence, the latent space of the Autoencoder loses its ability to extract physically meaningful patterns. In these cases, the method prioritizes minimizing the global reconstruction error, but the resulting modes do not provide additional insight into the underlying flow physics.

\begin{figure}[h]
    \centering
    \includegraphics[trim = 0 0 0 0, clip, width=\textwidth]{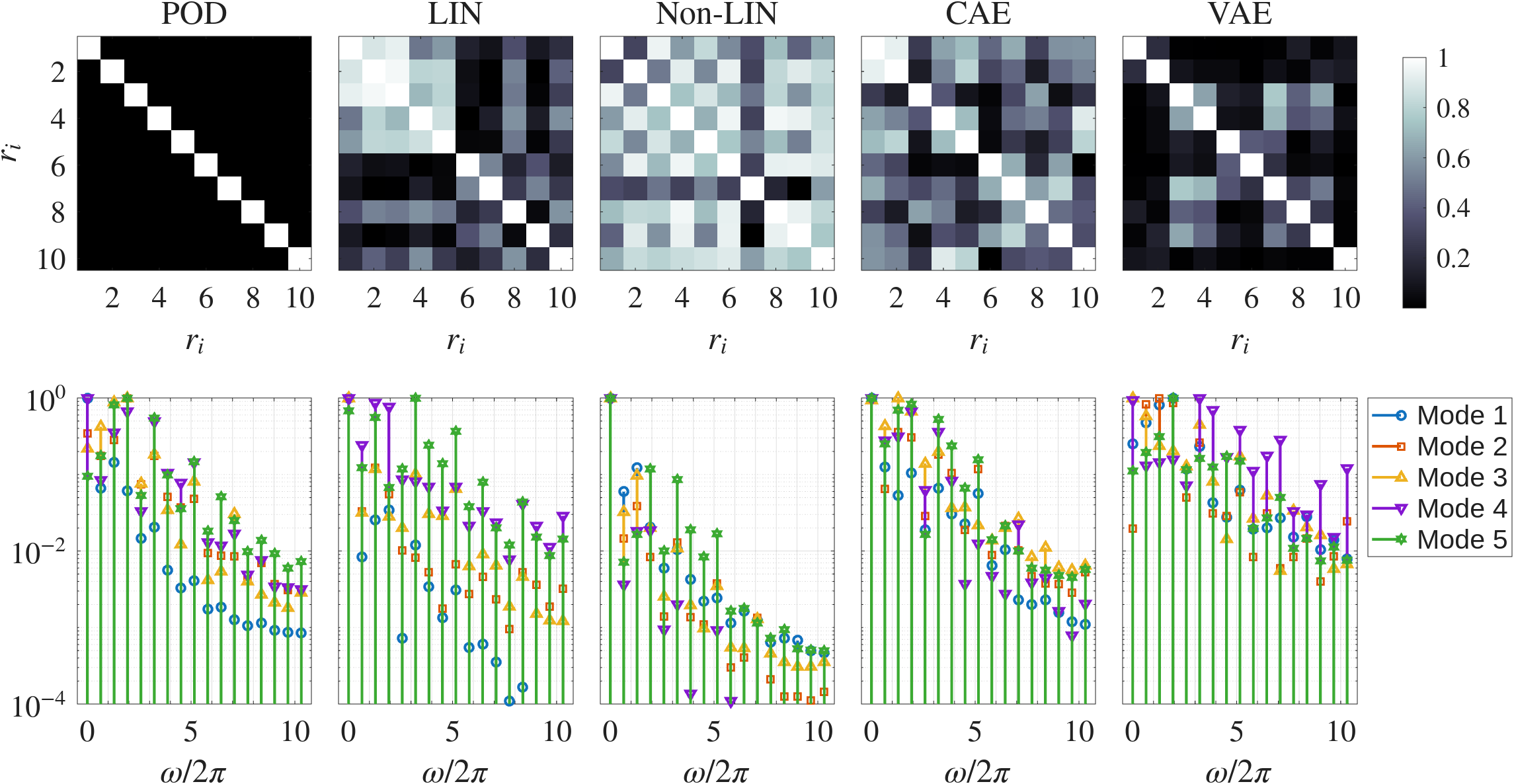}
    \caption{Same as Fig. \ref{fig:CorrelationMatrixFFT_r5} for $r=10$ corresponding to a cumulative energy of $E_{10} = 95.1\%$.}
    \label{fig:CorrelationMatrixFFT_r10}
\end{figure}

 Fig. \ref{fig:CorrelationMatrixFFT_r10} displays the correlation matrix and the Fourier frequency spectra associated to each of the configurations for $r=10$ corresponding to a cumulative energy of $E_{10} = 95.1\%$. A trend similar to the POD can be observed in the latent spaces of the VAE configuration. This configuration still produces a relatively large number of nearly orthogonal modes; however, as the latent dimension increases, the orthogonality gradually decreases, as reflected in Table~\ref{tab:RRMSE}, where the orthogonality measure drops from 0.87 to 0.73. This suggests that the network begins to represent small-scale, low-energy dynamics by introducing modes that contain redundant information. This effect is more pronounced in the Linear and Nonlinear Autoencoders, where many modes exhibit orthogonality values close to zero, indicating complete redundancy. The CAE behaves as an intermediate case, combining sets of orthogonal and non-orthogonal modes.  

Their behavior is consistent with that observed for the $r=5$ case, implying that the newly generated modes are linear combinations of the dominant ones. In other words, increasing the latent space dimension does not yield new independent dynamical features; instead, it reinforces existing flow structures already captured by the leading modes.

\begin{figure}[h]
    \centering
    \includegraphics[trim = 0 0 0 0, clip, width=\textwidth]{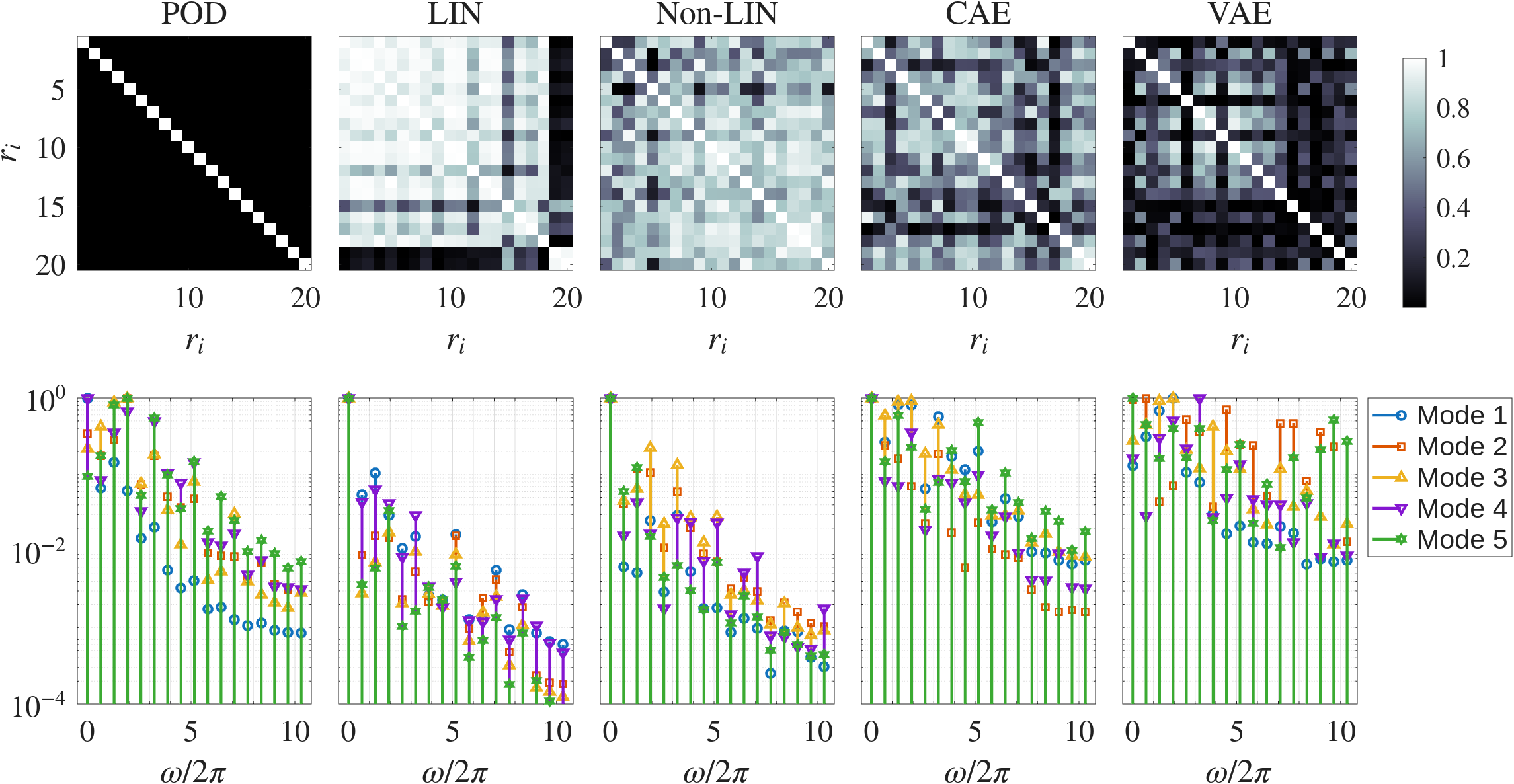}
    \caption{Same as Fig. \ref{fig:CorrelationMatrixFFT_r5} for $r=20$ corresponding to a cumulative energy of $E_{20} = 98.6\%$.}
    \label{fig:CorrelationMatrixFFT_r20}
\end{figure}

This behavior previously described becomes more pronounced as the latent dimension is increased to $r=20$ and $r=40$. Fig.~\ref{fig:CorrelationMatrixFFT_r20} displays the correlation matrices and FFT spectra for the $r=20$ case. For POD, the first twenty modes recover approximately $98\%$ of the total energy, suggesting that, in this case, modes beyond this threshold contribute mostly redundant information. Between modes ten and twenty, the singular values decrease by more than one order of magnitude relative to the leading mode. In these situations, the components tend to be associated with numerical noise, redundant flow features, or very small-scale dynamics of limited physical relevance.  

In the VAE, orthogonality progressively deteriorates, and some modes become fully correlated, exhibiting an orthogonality coefficient close to zero. The Linear Autoencoder performs worse in this respect: most of its modes are linearly dependent and thus repeated. The Nonlinear and Convolutional Autoencoders also identify only a small number of independent modes. In these cases, the latent representations reproduce the system dynamics in a redundant manner, optimizing the global reconstruction error but providing little additional insight into the underlying flow physics. This is also reflected in the orthogonality measure reported in Table~\ref{tab:RRMSE}, where the values for the Non-LIN configuration are very low and remain relatively modest for both the LIN and CAE cases.

\begin{figure}[h]
    \centering
    \includegraphics[trim = 0 0 0 0, clip, width=\textwidth]{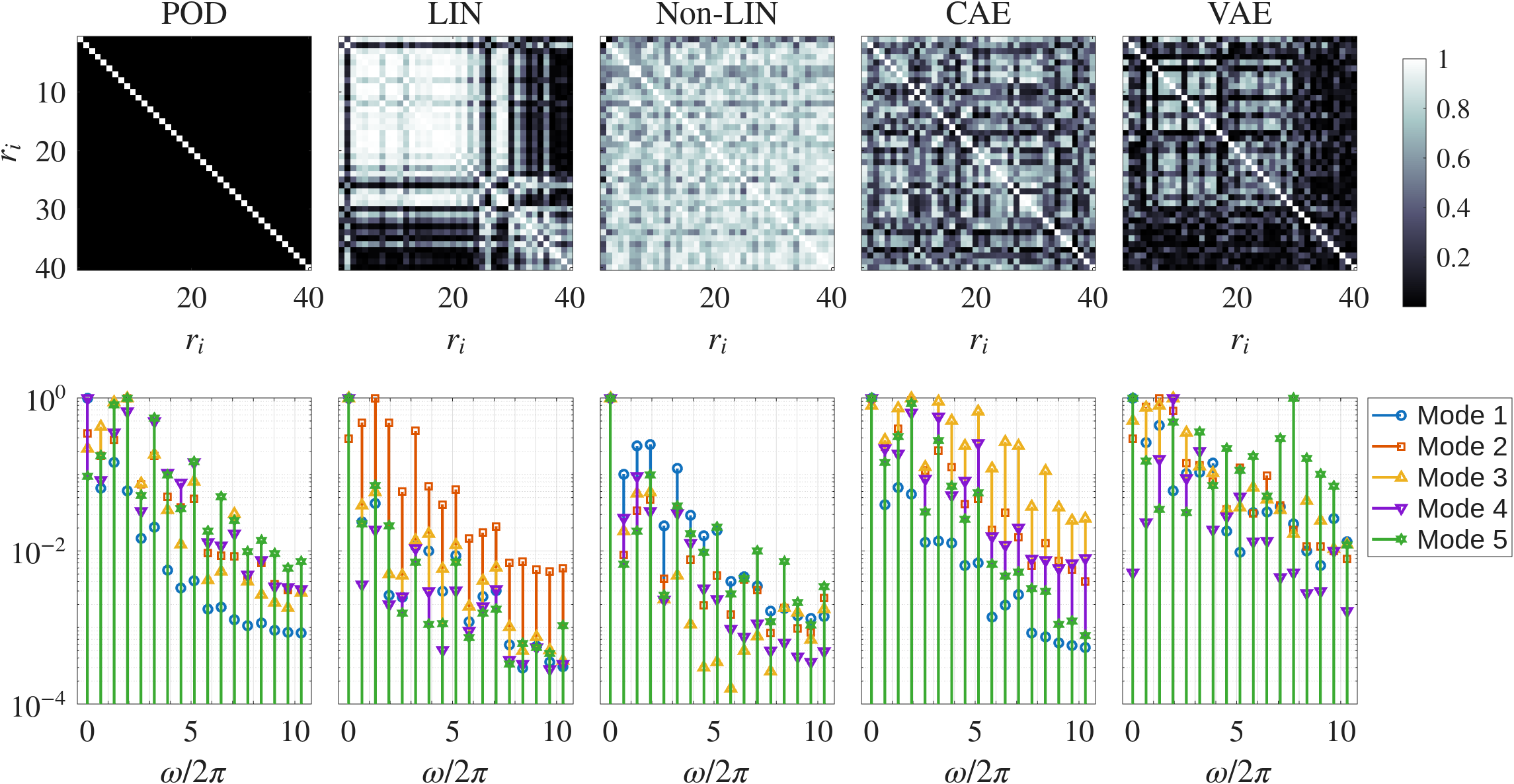}
    \caption{Same as Fig. \ref{fig:CorrelationMatrixFFT_r5} for $r=40$ corresponding to a cumulative energy of $E_{40} = 100\%$.}
    \label{fig:CorrelationMatrixFFT_r40}
\end{figure}

Finally, Fig.~\ref{fig:CorrelationMatrixFFT_r40} illustrates the case with latent space dimension $r=40$. Here, the reconstructed energy level reaches nearly $100\%$ for all methods, but the proportion of redundant modes increases substantially. The FFT spectra confirm this trend: the additional modes primarily reproduce the dominant frequency and its harmonics already captured by the leading modes. Consequently, enlarging the latent space beyond $r=20$ does not reveal new physical features, but instead amplifies redundancy within the modal representation.

\begin{figure}[h]
    \centering
    \includegraphics[trim = 0 0 0 0, clip, width=\textwidth]{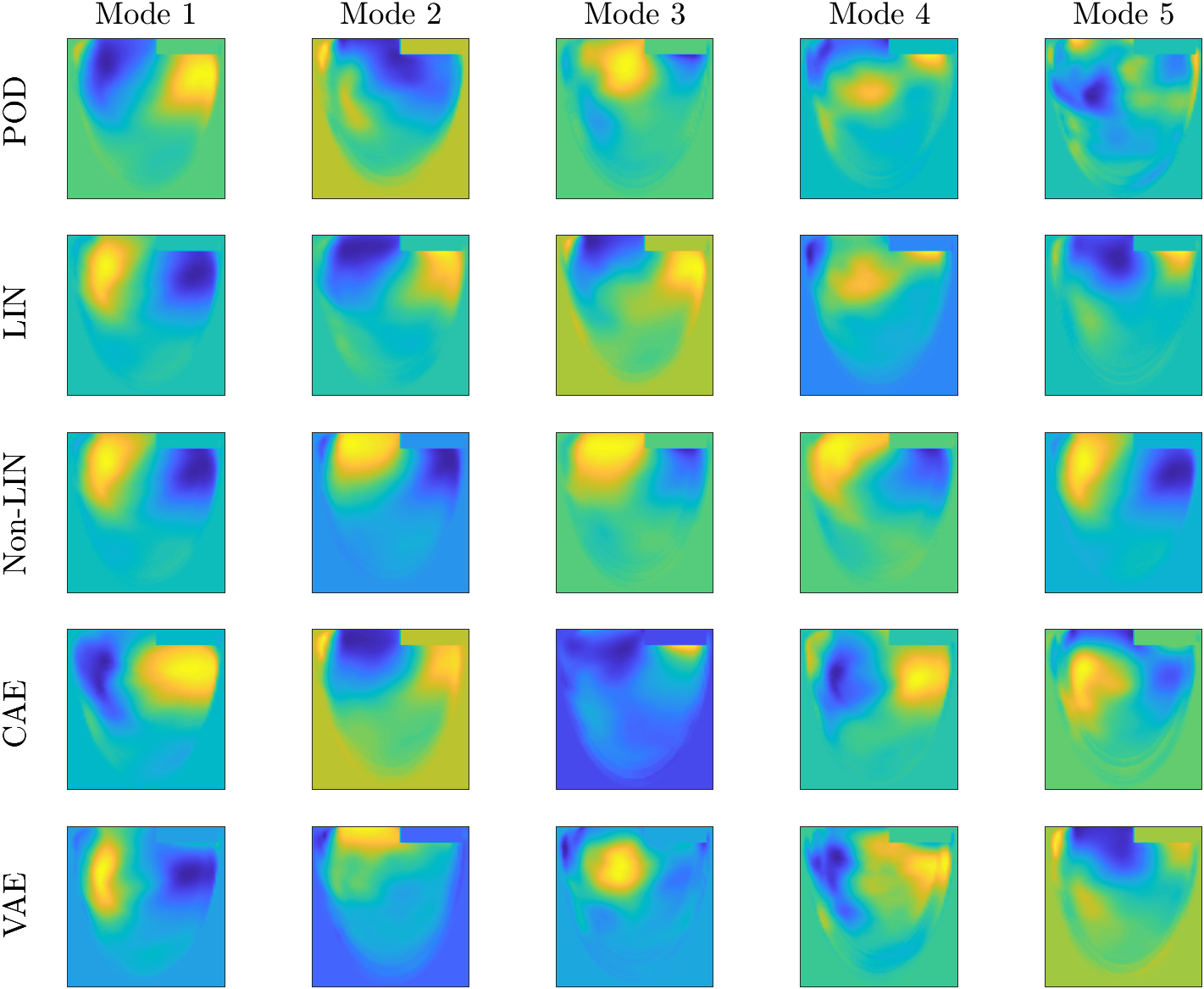}
    \caption{Spatial representation of the $v_z$ velocity component of the modes in the symmetry plane for $r=5$.}
    \label{fig:Modes2D}
\end{figure}

In addition to the global reconstruction level, we also inspect the spatial distribution of the $v_z$ velocity component corresponding to the modes extracted for $r=5$ in Fig.~\ref{fig:Modes2D}. The VAE modes exhibit a high degree of orthogonality, similar to the POD basis, despite being obtained through a nonlinear encoding process. In contrast, the LIN and Non-LIN configurations show higher linear dependence between modes, with several of them displaying repeated spatial structures even within the same latent dimension. This redundancy is consistent with the correlation and spectral analyses, indicating that these models tend to encode overlapping representations of the flow dynamics rather than isolating distinct physical features. The CAE configuration is again an intermediate case. Overall, while the POD and VAE approaches yield compact and interpretable modal bases, the other architectures primarily optimize reconstruction accuracy at the expense of physical interpretability.

%%%%%%%%%%%%%%%%%%%%%%%%%%%%%%%%%
%%%%%%%%%%%%%%%%%%%%%%%%%%%%%%%%%
\section{\label{sec:Conclusions} Conclusions}

The present analysis highlights the strong dependence of the Autoencoder performance on the chosen latent space dimension and architecture. For small encoding dimensions, both Proper Orthogonal Decomposition (POD) and Variational Autoencoders (VAE) recover physically interpretable modes that isolate coherent flow features. In these cases, the latent variables correspond to independent spatial structures with distinct frequency content, analogous to the classical modal decomposition. In this regime, Autoencoders are able to generate modal structures that qualitatively resemble POD modes, capturing the most energetic and coherent intraventricular flow features without introducing redundancy. However, as the latent dimension increases, redundancy arises in most Autoencoder configurations, leading to modes that are linearly dependent and often represent repeated flow patterns.

This loss of orthogonality reflects a transition from physically meaningful encodings to purely data-fitting representations, where the network prioritizes reconstruction accuracy over interpretability. Such degradation is also accompanied by the emergence of high-frequency or noise-like content in the AE modes; this in turn implies that the AE lose the ability to represent coherent flow physics and complicates their use in hybrid physics–ML formulations. Among all tested models, the VAE demonstrates the best compromise between reconstruction quality and physical interpretability, preserving orthogonal and complementary modes across different latent dimensions. Architectures with just fully connected layers, in contrast, tend to overfit the data and capture redundant spatial information when the latent space is not optimally constrained.
Beyond the methodological implications, these results hold particular relevance for cardiac applications, where coherent intraventricular flow patterns are strongly linked to ventricular performance, pathological remodeling, and early indicators of cardiac dysfunction. Interpretable modal structures are essential for building reduced-order models that can support clinical decision-making, accelerate patient-specific simulations, and enable real-time hemodynamic assessment in diagnostic or interventional scenarios. Ensuring that Autoencoders preserve physically meaningful patterns is therefore critical for their eventual integration into cardiovascular modeling pipelines.

Overall, the study underscores that the selection of the latent space dimension is critical for achieving a balance between compactness, reconstruction accuracy, and physical insight. When properly tuned, Autoencoders, particularly the VAE variant, can provide an interpretable and data-efficient alternative to traditional modal decompositions such as POD for analyzing complex fluid dynamics. These findings also emphasize the importance of preserving orthogonality and avoiding redundant representations when Autoencoders are intended for physics-informed reduced-order models, ensuring that the learned modes remain linked to the underlying flow mechanisms rather than to data-driven artifacts.

%%%%%%%%%%%%%%%%%%%%%%%%%%%%%%%%%
%%%%%%%%%%%%%%%%%%%%%%%%%%%%%%%%%
\section{Data Availability Statement}
The data that support the findings of this study are available upon reasonable request.
\\
Should the reader wish to gain a more detailed understanding of the process by which the databases were obtained, as well as other pertinent information, they are invited to visit the following website, where they will find the relevant codes and tutorials in Ref. \cite{ModelFLOWsCardiac}.

%%%%%%%%%%%%%%%%%%%%%%%%%%%%%%%%%
%%%%%%%%%%%%%%%%%%%%%%%%%%%%%%%%%
\section{Acknowledgements}
The authors acknowledge the grants TED2021- 129774B-C21 and PLEC2022-009235 funded by MCIN/AEI/ 10.13039/501100011033 and by the European Union “NextGenerationEU”/PRTR and the grant PID2023-147790OB-I00 funded by MCIU/AEI/10.13039/501100011033/FEDER, UE. The authors gratefully acknowledge the Universidad Politécnica de Madrid (www.upm.es) for providing computing resources on Magerit Supercomputer.

\bibliography{biblio}

\end{document}